\documentclass[twoside]{LCWS11}
\usepackage[latin1]{inputenc}
\usepackage[dvips]{graphicx,epsfig,color}
\usepackage{wrapfig,rotating}
\usepackage{amssymb,amsmath,array}
\usepackage{cite}
\usepackage{subfigure,epsfig}
 \newcommand{\GeV}{\:\mathrm{GeV}}
 \newcommand{\avg}[1]{\left< #1 \right>} % for average
 \newcommand{\mm}{\:\rm{mm}}
 \newcommand{\cmcube}{\:\rm{cm}^{3}}
  \newcommand{\MeV}{\:\mathrm{MeV}}
  \newcommand{\g}{\:\mathrm{g}}
  \newcommand{\m}{\:\rm{m}}
\begin{document}
\title{
%%%%   Paper title goes here  %%%%%%%%%%%%%%
Positrons sources and related activities for Future Linear Collider at LAL Orsay Laboratory} %% 
%***********************************************************************
% AUTHORS INFORMATION AREA
%***********************************************************************
\author{Olivier Dadoun and collaborators.
% Optional short acknowledgment: remove next line if non-needed
%\thanks{IPN Lyon France and BINP, Av. Lavrentyeva, 11, 630090 Novosibirsk, Russia }
% DO NOT MODIFY THE FOLLOWING '\vspace' ARGUMENT
\vspace{.3cm}\\
% Addresses and institutions (remove "1- " in case of a single institution)
1- LAL, IN2P3-CNRS and Universit\'e Paris Sud, 91898 Orsay Cedex, France\\
dadoun@lal.in2p3.fr
%% Remove the next three lines in case of a single institution
}
%%***********************************************************************
% END OF AUTHORS INFORMATION AREA
%***********************************************************************
\maketitle

%%%%%%%%%%%%%%%%%%%%
\begin{abstract}
In the context of the positrons sources studies for the Future Linear Collider, the 
Accelerator Department at LAL Orsay is involved since several years in different activities 
both experiments and simulations.
\end{abstract}
%%%%%%%%%%%%%%%%%%%%%%
\section{Introduction}
Basically the positron production requires an intense gamma flux (produced by a radiator) impinging on a target converter~\footnote{N.B.: in the case of conventional scheme the radiator and the converter are the same target.}. The positron created are then trapped and accelerated into a capture section located downstream the converter. \\
Two solutions using two different radiator are under study~:
\begin{enumerate}
\item the Compton scheme based on circular polarized laser (baseline for polarized positron production at CLIC and an alternative for  ILC)~;
\item the hybrid scheme based on a crystal (baseline for unpolarized positron production at CLIC and under studied for ILC).
\end{enumerate}
Concerning the capture section innovative concepts are under investigation.
%pair creation occur are produce using gamma pair production inside a target, the converter. The  production of gamma could be done using  two different kind of radiator~%%\footnote{NB: in the case of the conventional scheme the radiator and the converter are the same target.}. By a laser in the case of Compton scheme (polarized positron %production baseline for CLIC and an alternative method for  ILC)  or a crystal in the case of the hybrid scheme (unpolarized positron production baseline for CLIC). The %positron created need to be capture and accelerate by the capture section.
%The radiator~\footnote{NB: in the case of conventional scheme the radiator and the converter are the same target.} could be a laser (in the case of Compton scheme) or a %crystal in the case of the hybrid scheme production needs basically.  The positron created need to be capture and accelerate by the capture section.
%The high intensity and low emittance of the positron beam required by the Future Linear Collider at the interaction region
%needs to explore new concept of production as well as new concept of capture section.\\

%This paper reviews  the Accelerator Department LAL Orsay activities concerning~: those two kind of radiator  Compton and the hybrid positron source as well as  
%innovative concepts of the capture section located downstream the target positron production.

%%%%%%%%%%%%%%%%%%
\section{The Compton scheme}
\begin{wrapfigure}{r}{0.46\columnwidth}
\centerline{\includegraphics[width=0.39\columnwidth]{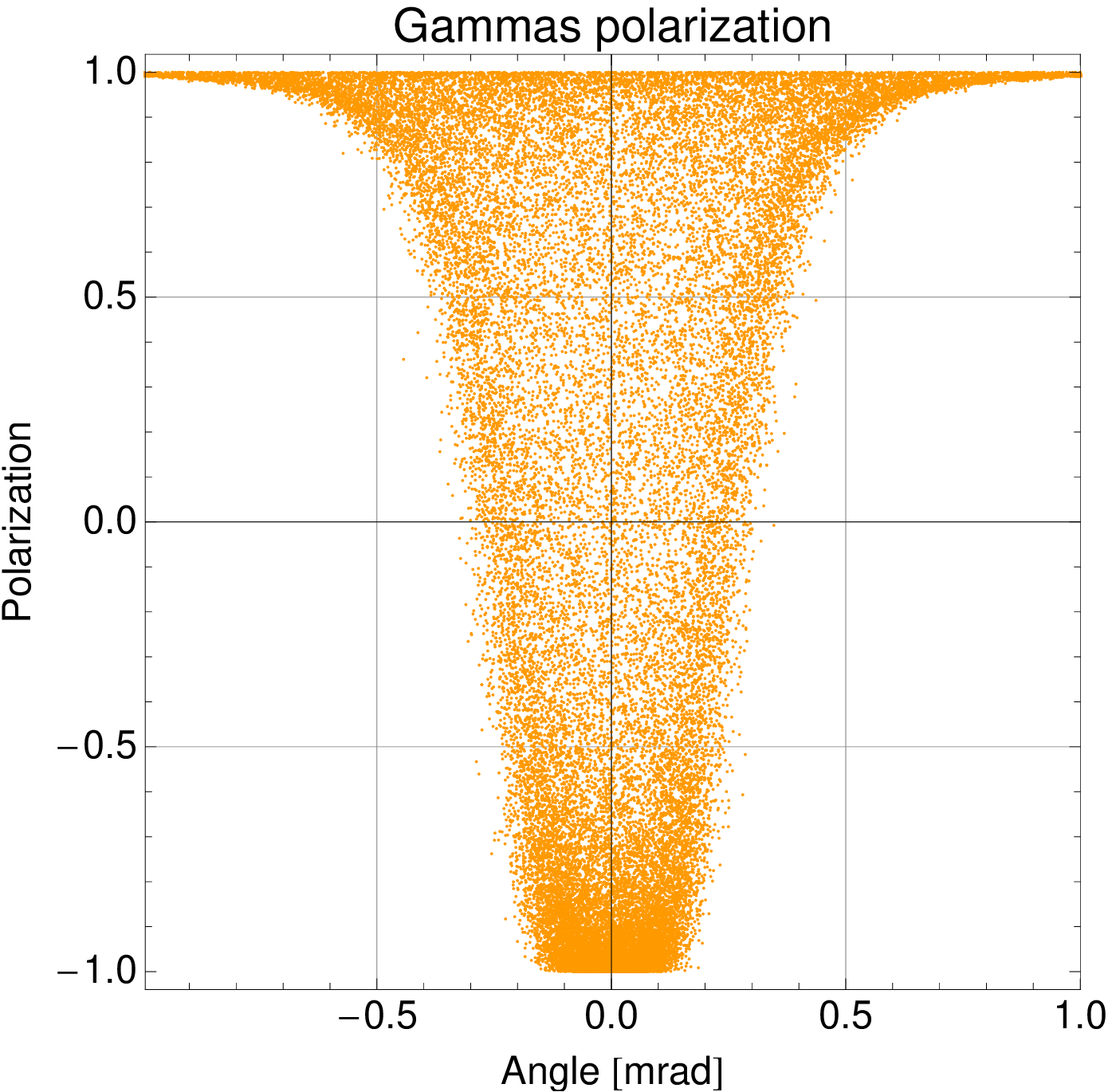}}
%\caption{Gamma polarization versus angle at the production.}
\label{fig:cain}
% Incident electron beam energy $1.3\GeV$ and laser photon energy $1.2\eV$.}\label{fig:cain}
\end{wrapfigure}
\subsection{Polarized positrons principle based on Compton scheme}
At Snowmass 2005, a polarized-positron source based on laser-Compton scattering was proposed for the ILC~\cite{snow mass}. 
Circularly polarized laser photons are backscattered by an electron beam and the Compton photons materialize into longitudinally polarized 
$\mathrm{e}^+\mathrm{e}^-$ pairs in a thin target.
%Circular polarized photons generated by laser-electron scattering are converted into linear polarized positrons via pair production in a target.
% A typical shape of gamma Compton scattering is shown in opposite figure.
The correlation polarization-angle of this process shown in the opposite figure, give us the ability to select photon polarization using appropriate collimators . This allows, then, to control  
the positron polarization production.

However the low cross section of the Compton scattering requires a high current electron beam
%\footnote{For positron source application a high energy electron should be considered, typically above $1\GeV$.} 
and high average power laser.  To achieve such power (in the order of Megawatt)  a Fabry-Perrot Cavity (FPC) is filled with a pulsed laser beam focused in a small waist allowing to reach a  high power at the interaction point of the cavity.  The use of a two mirrors cavity is unstable when one works in small mode waist. 
This is one of the reasons why four mirrors cavities is considered for Mighty laser experiment. 

\subsection{Mighty laser experiment at KEK}
\begin{figure} [htp]
\begin{center}	
 \subfigure[Layout of th ATF at KEK.]{
 \epsfig{file=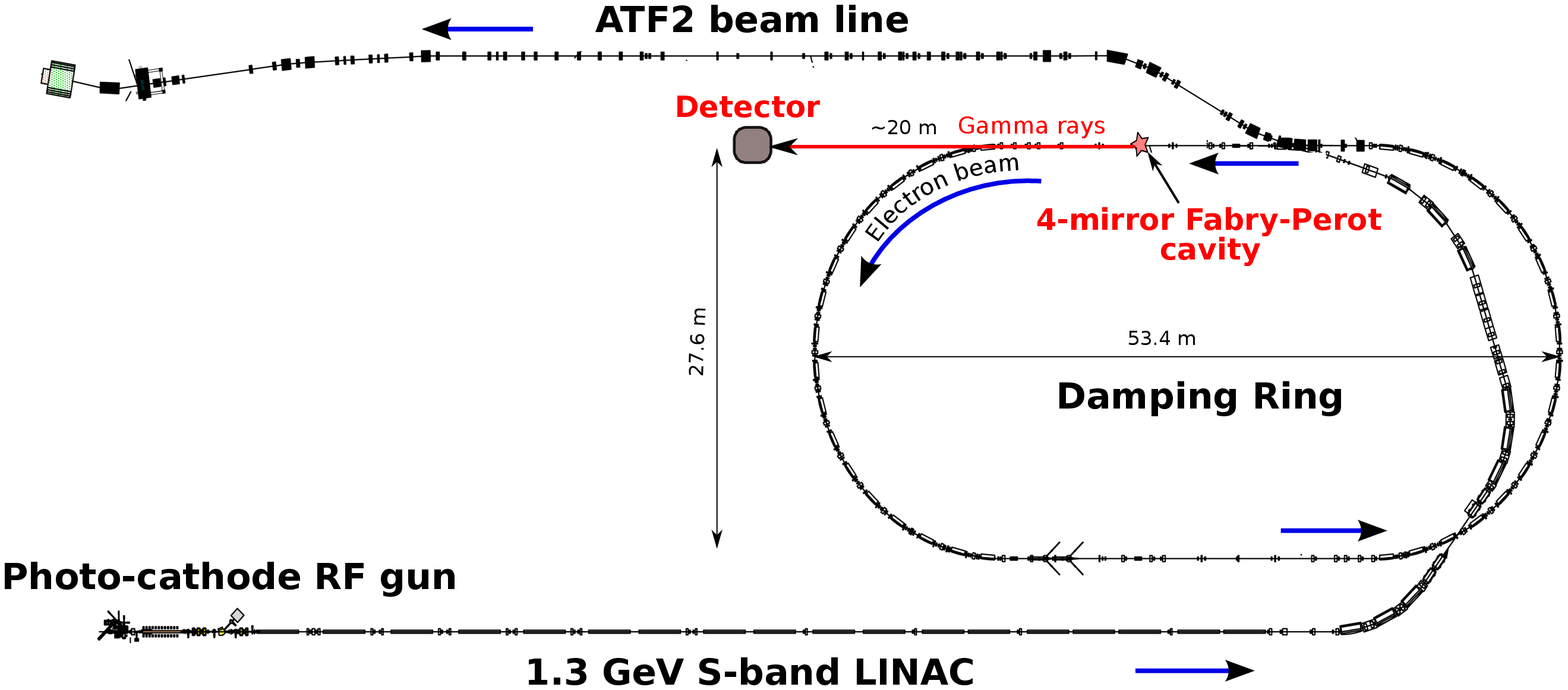,width=0.57\columnwidth}\label{fig:atf}}
  \subfigure[Drawing of the 4-mirror FPC. ]{
  \epsfig{file=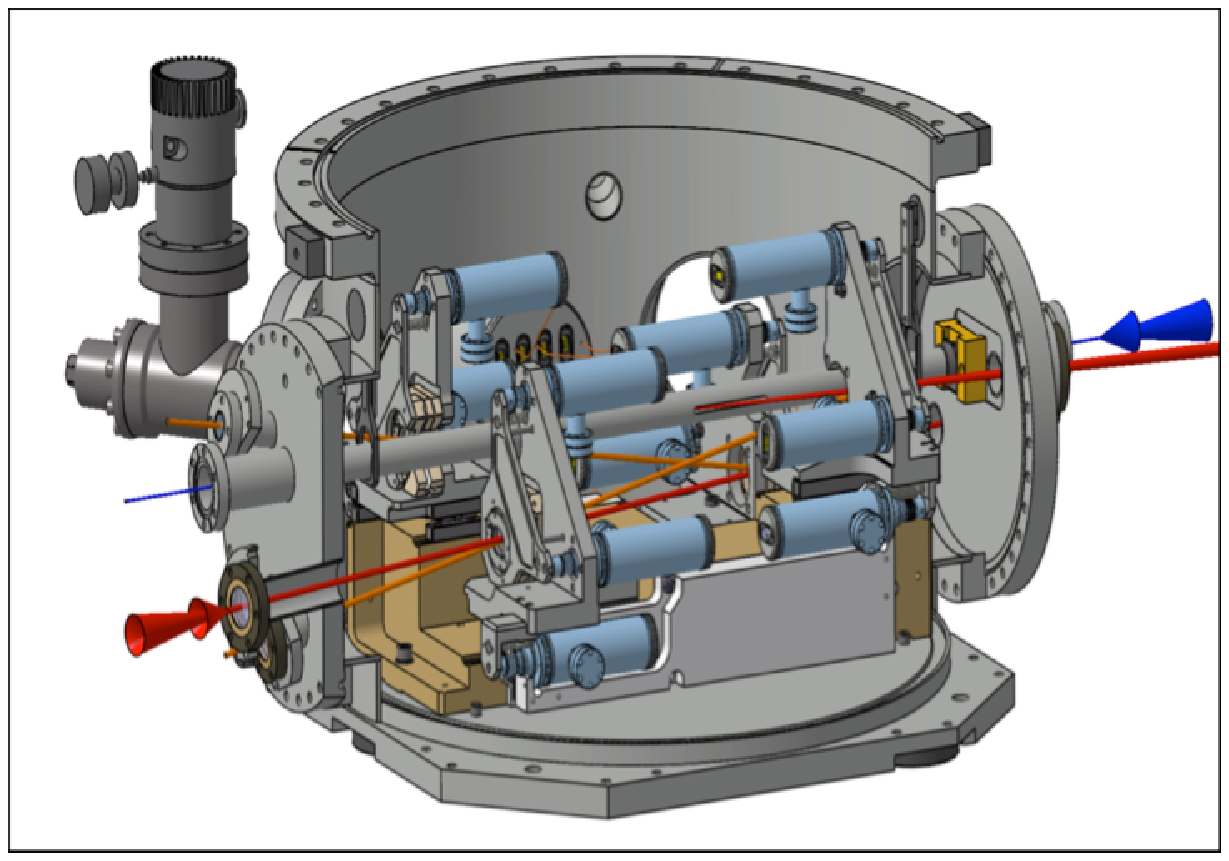,width=0.4\columnwidth}\label{fig:4mirror}}%
\caption{On the left, position of the 4-mirror FPC in the layout of the ATF at KEK.
On the right, drawing the FPC. The incoming laser beam is represented by the left red arrow and the incoming electron beam by the right blue arrow.}
\end{center}
\end{figure}

A prototype of such high finesse four-mirror FPC has been installed at the KEK ATF Figure~\ref{fig:atf} and is described in details in~\cite{Bonis:2011hi}.
This prototype aims to contribute to a global R\&D effort to reach the hundredth kW stored in such kind of cavity.
A schematic drawing of this four mirrors cavity  is shown Figure~\ref{fig:4mirror}.
The optical system has been commissioned during summer 2010 and electron-photon  collisions were observed on the first attempt in October 2010.
Data analysis of this experiment can be found in~\cite{Akagi:2011hj}.
Preliminary results are resumed on table~\ref{table:bestFlux} with respect to the electron beam pulse structure.

%\begin{table}[h]
%\centering
%\caption{Highest integrated gamma ray flux achieved.}
%\begin{tabular}{|c|c|c|c|c|c|}
%\hline
%\small  \begin{tabular}[x]{@{}c@{}} Electron \\pulse structure\end{tabular} &\small  \begin{tabular}[x]{@{}c@{}}Total intensity\\over 0.2~ms\end{tabular} & \small  \begin{tabular}%[x]{@{}c@{}}Energy deposited\\ over 0.2~ms\end{tabular}& \small  \begin{tabular}[x]{@{}c@{}} Integrated flux\\ over 0.2~ms\end{tabular}&  \small  \begin{tabular}[x]{@{}c@{}}  %Integrated flux\\ over 1~s\end{tabular} & \begin{tabular}[x]{@{}c@{}} \small Systematic \\ error \end{tabular} \\\hline \hline
%\small 1 train &\small 893~mV& \small 23750 MeV & \small 990~ $\gamma$ & \small $\sim4.9\times10^6~ \gamma$ & \small 7\% \\ \hline
%\small 2 trains &\small 910~mV & \small 24210 MeV& \small 1010~$\gamma$ & \small $\sim5.0\times10^6~ \gamma$& \small 7\%  \\ \hline
%\small 3 trains &\small 1010~mV & \small 26800 MeV & \small 1120~ $\gamma$ & \small $\sim5.6\times10^6~ \gamma$& \small 7\%  \\ \hline
%\end{tabular}
%\label{table:bestFlux}
%\end{table}
\begin{table}[h]
\centering
\caption{Highest integrated gamma ray flux achieved.}
\begin{tabular}{|c|c|c|c|}
\hline
\small  \begin{tabular}[x]{@{}c@{}} Electron \\pulse structure\end{tabular} & \small  \begin{tabular}[x]{@{}c@{}} Integrated flux\\ over 0.2~ms\end{tabular}&  \small  \begin{tabular}[x]{@{}c@{}}  Integrated flux\\ over 1~s\end{tabular} & \begin{tabular}[x]{@{}c@{}} \small Systematic \\ error \end{tabular} \\\hline \hline
\small 1 train  & \small 990~ $\gamma$ & \small $\sim4.9\times10^6~ \gamma$ & \small 7\% \\ \hline
\small 2 trains & \small 1010~$\gamma$ & \small $\sim5.0\times10^6~ \gamma$& \small 7\%  \\ \hline
\small 3 trains & \small 1120~ $\gamma$ & \small $\sim5.6\times10^6~ \gamma$& \small 7\%  \\ \hline
\end{tabular}
\label{table:bestFlux}
\end{table}

The MightyLaser project has demonstrated the production of gamma rays using a four mirror FPC. Next campaign to increase the number to gamma produced is under preparation.
%%%%%%%%%%%%%%%%%%%%%%%%%%%
\section{Hybrid scheme}
\subsection{Presentation}
CLIC considers as a baseline a method based on a combined crystal and amorphous tungsten targets~: so-called the hybrid source~\cite{artru}. A  $5\GeV$ electron beam impinges on a tungsten crystal oriented on its $\avg{111}$ axis~\cite{oim1}. To limit the energy deposition in the amorphous target, the
 charged particles are swept off after the crystal. Only the photon beam impinges on the amorphous target. In the goal of decreasing the energy deposition density in the converter some studies on the target geometry have been done. Concerning the simulation of the interaction of fast charged electron with the crystal a Geant4 event generator have been implemented with the partnership of IPN Lyon (France).
%%%%%%%%%%%%%%%%%%%%%%%%%%
 \subsection{Converter study : granular amorphous target}
The very intense incident electron beam considered in the linear collider projects requires resistant targets.  Both concerning the total energy and the  energy density deposited in the target converter.  For the latter the Peak Energy Deposition Density (PEDD) is a critical parameter.
In~\cite{xu-chehab} the authors, using the experience of the target for muon collider,  shown that using a granular target instead of compact reduces the PEDD (Table~\ref{tab:granular}).

\begin{table}[h!]
\begin{center}
\begin{tabular}{|c|c|c|c|c|c|c|c|}
\hline
&\small t($\mm$)& \small$\mathrm{e}^+/\mathrm{e}^-$ & \small  \begin{tabular}[x]{@{}c@{}} PEDD\\ $(\GeV/\cmcube/\mathrm{e}^-)$\end{tabular}& 
\small  \begin{tabular}[x]{@{}c@{}} $\Delta\mathrm{E}_{dep}$\\$(\MeV/\mathrm{e}^-)$\end{tabular}&\small N-Layers &\small N-Spheres &\small $\rho(\g/\cmcube$)\\
\hline
\small Compact & \small 8.00 & \small 13.30 & \small 2.24 & \small 523 &  & &\small 19.3\\
\hline
\begin{tabular}[x]{@{}c@{}}\small Granular\\$\small \mathrm{r}=1.0\mm$\end{tabular} &  \small 10.16 & \small 12.50 &\small 1.80&\small 446&\small 3&\small 864&\small 13.9\\
\hline
\begin{tabular}[x]{@{}c@{}}\small Granular\\$\small \mathrm{r}=0.5\mm$\end{tabular} &  \small 11.60 & \small 13.45 & \small 2.33 & \small 613 & \small 7 & \small 8064 & \small 13.9 \\
\hline
\end{tabular}
\caption{Comparison of compact and granular target.}
\label{tab:granular}
\end{center}
\end{table}
The high ratio of surface volume of the spheres ($3/\mathrm{r}$) makes easier the thermal dissipation. In the scheme considered for the converter of the hybrid source, the spheres are arranged in staggered rows. The choice of the diameter of the spheres and of  the number of rows is related to two quantities~: the required positron yield and the PEDD.
A comparison between two granular targets and a compact one giving close yield values is presented on Table~\ref{tab:granular}.

\subsection{Simulation}
The radiation emitted from the interaction between a fast electron  and the the individual atoms of  an amorphous media is called incoherent bremsstrahlung.
In crystalline targets and when the electron trajectory is close enough to a major crystallographic axis the amplitudes of the bremsstrahlung emission at certain wavelengths may interfere constructively.
This will result in an enhancement of the intensity at this wavelength compared to ordinary incoherent emission.  
\begin{figure} [htp]
\centering %
\includegraphics[width=0.65\columnwidth]{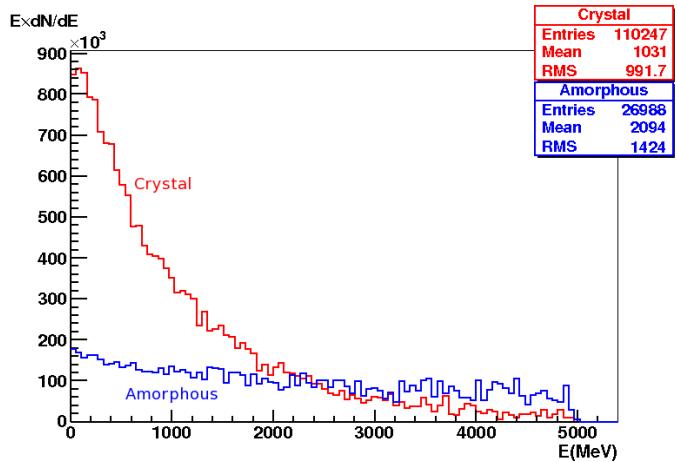}%
\caption{Emerging photons from 6000 electrons at $5\GeV$ impinging on $1.4\mm$ of tungsten crystal oriented on its  $\avg{1 1 1}$  axis and amorphous target. }
\label{fig:crystalAmorph}
\end{figure}
This type of radiation is called coherent bremsstrahlung. Moreover if particles enter with glancing angle to axis they may be trapped in "channels" performing stable periodical trajectories along atoms rows or planes.
This radiation is then called channeling radiation or Kumakhov radiation.  
Those physics processes have been implemented as an event generator for  Geant4~\cite{oim2}. Comparison of a amorphous and a crystal tungsten target with the same thickness are shown Figure~\ref{fig:crystalAmorph}.  The large number of photon are used to impinge on a converter $2\m$ downstream the radiator~\cite{oim1}.

\section{Positron Capture pre-accelerating section}
\begin{wrapfigure}{r}{0.5\columnwidth}
\centerline{\includegraphics[angle=90,width=0.5\columnwidth]{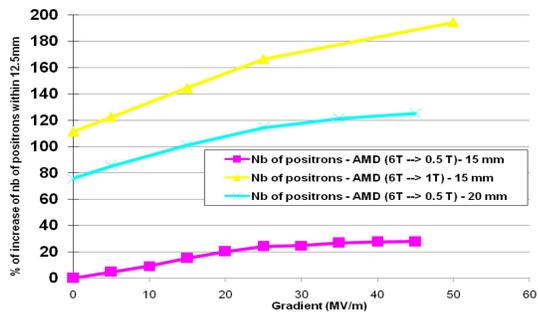}}
\caption{Positron increase versus the gradient.}
\label{fig:amd-studies}
\end{wrapfigure}
Due to the huge amount of multiple Coulomb scattering inside the converter, the positron beam has a large emittance and consequently is difficult to transport. Before being injected into to the damping ring to decrease their emittances the positrons should be trapped into specific magnetic device.
The capture section is composed of an Adiabatic Matching Device (AMD) and a pre-accelerator encapsulated in a solenoid. The AMD has a large energy acceptance and hence increases the number of accepted positrons. 
%\begin{figure} [htp]
%\centering %
%\includegraphics[angle=90,width=0.6\columnwidth]{fig/AMDstudy.eps}%
%\caption{Positron increase versus the gradient}
%\label{fig:amd-studies}
%\end{figure}
%Basically the AMD uses a slowly decreasing magnetic field changing the particle's transverse momentum. 
At the end of the AMD the magnetic field is equal to the magnetic field solenoid which encapsulates the pre-accelerator.  
The studies for using a standing wave 4-cells cavity to accelerate the positron within the AMD have been investigated~\cite{freddy}, the results of different 
senarios are shown Figure~\ref{fig:amd-studies}. The figures indicates the percentage 
of increase of the number of positrons as a function of the cavity gradient within the AMD field and considering the different maximum and minimum magnetic field.

Concerning the studies of the pre-accelerator itself two modes of operation could be applied. The first one is based on acceleration of the positrons straight after the AMD.  
This method have been already used  in different collider.  The second one requiere using the first accelerating structure in 
a decelerating mode such that a large number of positrons is captured~\cite{freddy}. 

%The different shapes of the energy distribution versus longitudinal positrons for the accelerating mode is presented on Figure~\ref{fig:acc:decc}.
%\begin{figure} [htp]
%\begin{center}	
 %\subfigure[Acceleration scenario.]{
 %\epsfig{file=fig/acc_scenario.eps,width=0.44\columnwidth}\label{fig:acc}}
  %\subfigure[Decceleration scenario.]{
  %\epsfig{file=fig/decc_scenario.eps,width=0.45\columnwidth}\label{fig:decc}}%
%\caption{Energy distribution versus longitudinal position for the different scenario.}
%\label{fig:acc:decc}
%\end{center}
%\end{figure}
The capture section studies have shown that the required number of positron at the end of the pre-accelerator  for CLIC is matching the requirements.

\section{Conclusion}
%% section headers !
The Accelerator Department at LAL Orsay has works in different aspect of the positron production for the future linear collider, from the production up to the
capture section. Two theses are under preparation and will be submitted before the end of 2012.
\section{Acknowledgments}
The collaborators include X. Artru \& R. Chehab (IPN Lyon France) and V. Strakhovenko (BINP Novosibirsk Russia).
% (if I mentionned them here I will exceed the  number of page requested). 
% ****************************************************************************
% BIBLIOGRAPHY AREA
% ****************************************************************************

\begin{footnotesize}
% IF YOU DO NOT USE BIBTEX, USE THE FOLLOWING SAMPLE SCHEME FOR THE REFERENCES
% ----------------------------------------------------------------------------

\end{footnotesize}

% ****************************************************************************
% END OF BIBLIOGRAPHY AREA
% ****************************************************************************

\end{document}